# Evolutionary mismatch and the role of GxE interactions in human disease

Amanda J. Lea*, Andrew G. Clark, Andrew W. Dahl, Orrin Devinsky, Angela R. Garcia, Christopher D. Golden, Joseph Kamau, Thomas S. Kraft, Yvonne A. L. Lim, Dino Martins, Donald Mogoi, Paivi Pajukanta, George Perry, Herman Pontzer, Benjamin C. Trumble, Samuel S. Urlacher, Vivek V. Venkataraman, Ian J. Wallace, Michael Gurven[†], Daniel Lieberman[†], Julien F. Ayroles*

[†]These authors contributed equally
*Corresponding authors' e-mail: amanda.j.lea@vanderbilt.edu and jayroles@princeton.edu

AL - Department of Biological Sciences, Vanderbilt University, Nashville, TN, USA
AL - Child and Brain Development, Canadian Institute for Advanced Research, Toronto, Canada
AC - Department of Computational Biology, Cornell University, Ithaca, NY, USA
AC - Department of Molecular Biology and Genetics, Cornell University, Ithaca, NY, USA
AD - Section of Genetic Medicine, University of Chicago, Chicago, IL, USA
OD - Department of Neurology, NYU Langone Medical Center, New York, NY, USA
OD - Comprehensive Epilepsy Center, NYU Langone Medical Center, New York, NY, USA
AG - Center for Evolution and Medicine, Arizona State University, Tempe, United States
CG - Department of Nutrition, Harvard T.H. Chan School of Public Health, Boston, MA, USA
JK - Department of Biochemistry, School of Medicine, University of Nairobi, Nairobi, Kenya
JK - Institute of Primate Research, National Museums of Kenya, Nairobi, Kenya
TK - Department of Anthropology, University of Utah, Salt Lake City, USA
YL - Department of Parasitology, Faculty of Medicine, Universiti Malaya, Kuala Lumpur, Malaysia
DM1 - Turkana Basin Research Institute, Turkana, Kenya
DM1 - Department of Ecology and Evolution, Princeton University, Princeton, NJ, USA
DM2 - Director at County Government of Laikipia, Nanyuki, Kenya
PP - Department of Human Genetics, David Geffen School of Medicine at UCLA, Los Angeles, CA, USA
PP - Institute for Precision Health, David Geffen School of Medicine at UCLA, Los Angeles, CA, USA
GP - Department of Anthropology, Pennsylvania State University, University Park, PA, USA
GP - Department of Biology, Pennsylvania State University, University Park, PA, USA
GP - Huck Institutes of the Life Sciences, Pennsylvania State University, University Park, PA, USA
HP - Evolutionary Anthropology, Duke University, Durham, NC, USA
HP - Duke Global Health Institute, Duke University, Durham, NC, USA
BT - School of Human Evolution and Social Change, Arizona State University, Tempe, US
BT - Center for Evolution and Medicine, Arizona State University, Tempe, United States
SU - Department of Anthropology, Baylor University, Waco, TX, USA
SU - Child and Brain Development, Canadian Institute for Advanced Research, Toronto, Canada
VV - Department of Anthropology and Archaeology, University of Calgary, Calgary, Canada
IW - Department of Anthropology, University of New Mexico, Albuquerque, USA
MG - Department of Anthropology, University of California: Santa Barbara, Santa Barbara, CA, USA
DL - Department of Human Evolutionary Biology, Harvard University, Cambridge, MA, USA
JA - Department of Ecology and Evolution, Princeton University, Princeton, NJ, USA
JA - Lewis Sigler Institute for Integrative Genomics, Princeton University, Princeton, NJ, USA

**Abstract**

Globally, we are witnessing the rise of complex, non-communicable diseases (NCDs) related to changes in our daily environments. Obesity, asthma, cardiovascular disease, and type 2 diabetes are part of a long list of "lifestyle" diseases that were rare throughout human history but are now common. A key idea from anthropology and evolutionary biology—the evolutionary mismatch hypothesis—seeks to explain this phenomenon. It posits that humans evolved in environments that radically differ from the ones experienced by most people today, and thus traits that were advantageous in past environments may now be "mismatched" and disease-causing. This hypothesis is, at its core, a genetic one: it predicts that loci with a history of selection will exhibit "genotype by environment" (GxE) interactions and have differential health effects in ancestral versus modern environments. Here, we discuss how this concept could be leveraged to uncover the genetic architecture of NCDs in a principled way. Specifically, we advocate for partnering with small-scale, subsistence-level groups that are currently transitioning from environments that are arguably more "matched" with their recent evolutionary history to those that are more "mismatched". These populations provide diverse genetic backgrounds as well as the needed levels and types of environmental variation necessary for mapping GxE interactions in an explicit mismatch framework. Such work would make important contributions to our understanding of environmental and genetic risk factors for NCDs across diverse ancestries and sociocultural contexts.

**Introduction**

Non-communicable diseases (NCDs) such as cardiovascular disease (CVD), type II diabetes, and Alzheimer's are among the leading causes of death worldwide (Figure 1). NCDs are often difficult to prevent and treat, because they result from complex and poorly understood interactions between a person's genetic makeup and their environment. For example, cardiovascular disease (CVD) has a heritability of 40-50%, with dozens of loci now mapped through genome-wide association studies (1–3). However, when tallied together in an additive framework, these loci explain only a small fraction of the heritable genetic effect. This has led many to conclude that environmental risk factors—such as a diet high in processed foods and low levels of physical activity—interact with genetic variation to shape NCD risk (4, 5). In other words, genetic variation may predispose individuals toward physiological sensitivity or resilience in the face of environmental perturbations, a phenomenon known as "genotype x environment" (GxE) interactions (Box 1).

Despite major interest in GxE interactions in the context of NCDs, scientists have struggled in practice to identify them. There are many complex reasons for this, including that the relevant environmental factors are often unknown, difficult to measure, or minimally variable within the study population (e.g., most individuals in high income countries (HICs) consume processed foods). Further, large sample sizes are needed to test for interaction effects, and even more so to overcome the multiple testing burden incurred by testing for interactions between many genetic variants and many environments (6, 7). To overcome power issues, current state-of-the-art approaches have leveraged very large studies such as the UK Biobank to scan for interactions between genome-wide genetic variation and key lifestyle factors (e.g., smoking, diet, or physical activity) (8–11). However, these studies have not delivered as expected, and have only uncovered a handful of GxE interactions for NCDs like obesity, type II diabetes, and depression.

Here, we argue for a complementary approach informed by anthropological traditions, genomic tools, and evolutionary theory. In particular, we believe there is much to learn by 1) viewing GxE interactions through the lens of the "evolutionary mismatch" hypothesis and 2) partnering with genetically and environmentally diverse small-scale, subsistence-level populations to map them. The evolutionary mismatch hypothesis posits that traits that evolved under past selection regimes are often imperfectly or inadequately suited to modern environments, leading to "mismatches" in the form of NCDs (12–15). At the genetic level, we would thus expect that previously neutral or beneficial alleles are now disease-causing.

While we cannot go back in time to evaluate genotype-phenotype relationships in past environments, we can collaborate with populations that practice non-industrial, subsistence-level lifestyles that are arguably more "matched" to their recent evolutionary history (though we caution that, of course, no modern population is perfectly representative of ancestral conditions). Importantly, many subsistence-level populations are currently exposed to globalizing forces causing rapid environmental shifts; this situation creates a quasi-natural experiment for studying the transition from traditional to modern lifeways within a single group (16) (Figure 2A). Additionally, many subsistence-level groups have already been well-characterized ecologically and phenotypically through long-term work with anthropologists (Figure 2B; Box 2), setting the stage for integration of genomic studies.

In this Consensus, we argue that uniting an evolutionary mismatch framework, long-term anthropological work with subsistence-level groups, and cutting-edge genomic tools can increase our power to identify and understand GxE interactions. Specifically, because the mismatch framework provides clear expectations for the types of loci and environments we expect to affect NCDs, we can narrow the search space considerably. Further, by focusing on populations where Western diets and lifestyles are the exception rather than the norm, we can design studies that explicitly sample environmental extremes, thereby boosting power. Finally, by studying many genetically distinct populations under a uniting intellectual framework, we can identify new loci that have so far been invisible to studies focused on individuals of European descent. With these goals in mind, we first review the evolutionary mismatch hypothesis and discuss its current support at the phenotypic and genetic levels. Second, we propose consensus recommendations for integrating mismatch principles with molecular and genomic techniques, focusing on collaborations with subsistence-level groups. Third, we discuss the payoffs for scientists and study communities that would come from implementing these partnerships.

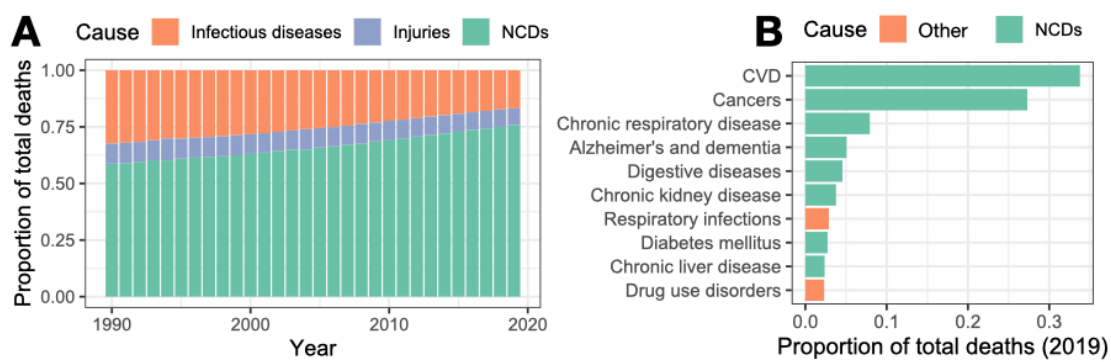

**Figure 1. Non-communicable diseases are the leading cause of death worldwide.** A) Proportion of worldwide deaths attributable to non-communicable diseases, communicable (infectious) diseases, and injuries through time. B) Proportion of deaths within the US in 2019, broken down by the top 10 causes of death. NCDs are highlighted in green. For both panels, data were sourced from ourworldindata.org and represent all ages.

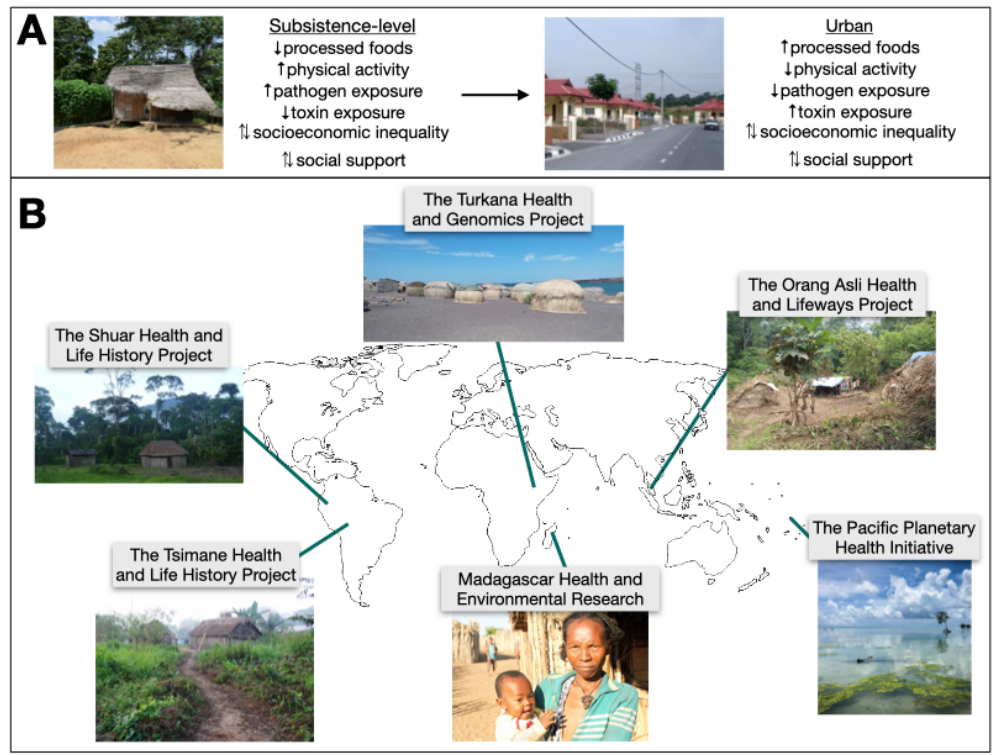

**Figure 2. Subsistence-level groups experiencing lifestyle change are a potential model for uncovering GxE interactions.** A) Subsistence-level groups faced with urbanization, market-integration, and modernization experience extreme variation in diet and physical activity levels, pathogen and toxin exposures, and social conditions. This list of environmental components for which there is extreme variation is not exhaustive, and in many cases will also be population specific. We highlight a few broad categories that tend to change consistently during lifestyle transitions. B) Studies such as The Turkana Health and Genomics Project (17, 18), The Orang Asli Health and Lifeways Project (19), The Pacific Planetary Health Initiative, Madagascar Health and Environmental Research (20–22), The Tsimane Health and Life History Project (23), and The Shuar Health and Life History Project (24, 25) all combine anthropological and biomedical data collection in transitioning societies, and are thus poised to uncover GxE interactions in the context of evolutionary mismatch. We note that this list is meant to be illustrative and only includes projects directed by authors of this Consensus; it does not by any means cover all ongoing projects of small-scale, subsistence-level groups.

**Overview of the evolutionary mismatch hypothesis**

An evolutionary mismatch is a condition that is more common or severe in an organism because it is imperfectly or inadequately adapted to a novel environment (26). While mismatches are not unique to humans, their frequency may be unusually high in our species. This is because human culture can generate rapid and profound environmental change: in just a few generations, industrialization has transformed human diets, physical activity patterns, and toxin exposure landscapes, especially in HICs, and these changes presumably contribute to the long list of NCDs that used to be rare or nonexistent (27–29).

For at least a century, a wide range of conditions have been assumed to be "diseases of civilization" or "lifestyle diseases" (30, 31), but mismatches need to be explicitly and rigorously tested according to three criteria (32). First, a mismatch condition should be more common or severe in the "novel" (e.g., post-industrial, HICs) relative to the "ancestral" environment (Figure 3A). Small-scale, subsistence-level societies typically stand in as the best available, though often imperfect, proxy for the "ancestral" condition in humans; this is because they experience a closer "match" between their recent evolutionary history and their current environments relative to individuals in HICs, though we caution they are not themselves "ancestral" populations.

In addition to the hypothesized mismatch condition being more prevalent in post-industrial versus subsistence-level groups, the second criteria is that it should also be tied to some environmental variable that differs between these groups (Figure 3B). One complication for achieving this is that NCDs arise from complex multifactorial causes, and thus, while between-population comparisons are necessary, they can be confounded by many covariates that must also be taken into account (e.g., sanitation, access to medical care, age structure).

The third criteria is that it is necessary to establish a molecular or physiological mechanism by which the environmental shift generates the proposed mismatch condition. At the genetic level, this should manifest as a locus for which 1) a variant exhibits a past history of positive selection and is associated with health benefits in the ancestral environment but health detriments in the novel environment or 2) past stabilizing selection has created a situation where two intermediate alleles have similar fitness in the ancestral environment, but one allele becomes associated with health detriments in the novel environment (Figure 3C; see also Box 1).

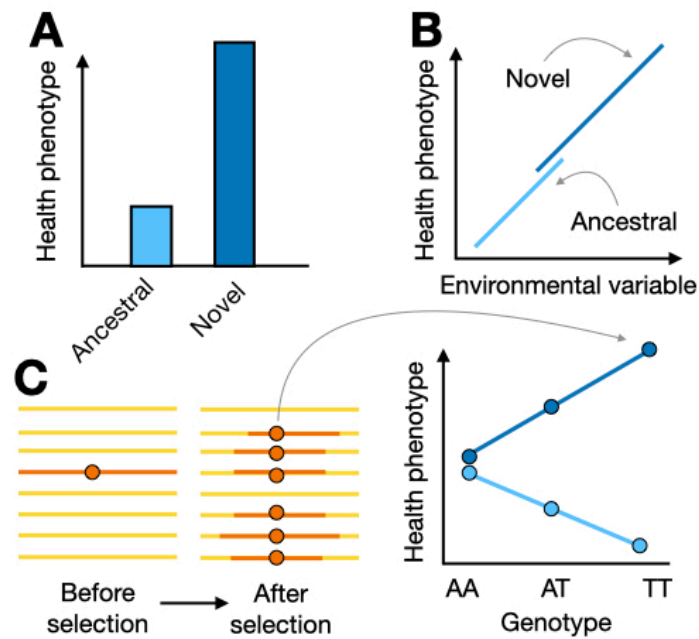

**Figure 3. Mismatch diseases must be tested according to three criteria**. A) Health phenotypes related to the hypothesized mismatch disease must be more common or severe in the novel versus ancestral environment. B) These health phenotypes must be attributable to an environmental variable, which will most often differ in mean and range between groups (e.g., physical activity influences cardiovascular health and is consistently higher in subsistence-level groups relative to HICs). C) It is necessary to establish a mechanism by which an environmental shift generates health issues. At the genetic level, this could manifest as a locus for which a variant exhibits a past history of positive selection and is associated with health benefits in the ancestral environment but health detriments in the novel environment. In panel C, horizontal lines represent haplotypes and the dark orange circle represents the selected variant. In all panels, dark blue represents the novel environment and light blue represents the ancestral environment.

**Current evidence for evolutionary mismatch at the phenotypic level**

Scientists have been relatively successful at testing the first two criteria for mismatch, especially in the context of CVD, the single largest cause of mortality worldwide (33). In support of the first criteria, subsistence-level groups experience remarkably low rates of CVD (29, 34, 35) relative to HICs, as well as minimal age-associated increases in CVD or its biomarkers (e.g., hypertension, cholesterol) (36–38) (Figure 4A). Studies of small-scale societies in the midst of socioeconomic transition have demonstrated within-population effects of industrialization (17, 39, 40), strengthening the findings from between-population comparisons.

In support of the second criteria, recent work has also isolated salient environmental changes by which industrialization promotes CVD. People in subsistence-level communities are generally very physically active, accruing 5-10 times more daily physical activity than adults in Europe, the U.S., and other HICs (41, 42). Moderate to vigorous physical activity increases cardiac output promoting nitric oxide production and arterial elasticity (43, 44), it also decreases baseline levels of inflammation, which plays a critical role in all aspects of CVD (45). Within industrialized populations, individuals accruing daily physical activity similar to those of subsistence-level individuals experience similarly low rates of CVD as well as NCD-related mortality (46) (Figure 4B). However, while physical activity plays a critical role in averting CVD, it is not a panacea and several other factors are surely important. For example, relative to HICs, subsistence-level groups subsist on diets dominated by unprocessed or minimally processed foods and experience different types and degrees of social integration and inequality—all of which impact CVD risk (47–49).

Finally, we note that while we have focused this section on CVD as an illustrative example of the type of comprehensive evidence required for diagnosing a mismatch disease, several other conditions also have relatively clear evidence for the first two criteria for mismatch. For example, inflammatory and autoimmune disorders have increased during the twentieth century, which has been linked to a reduced exposure to parasites and microorganisms (a phenomenon attributed to the "hygiene hypothesis" or "old friends hypothesis") (50–52).

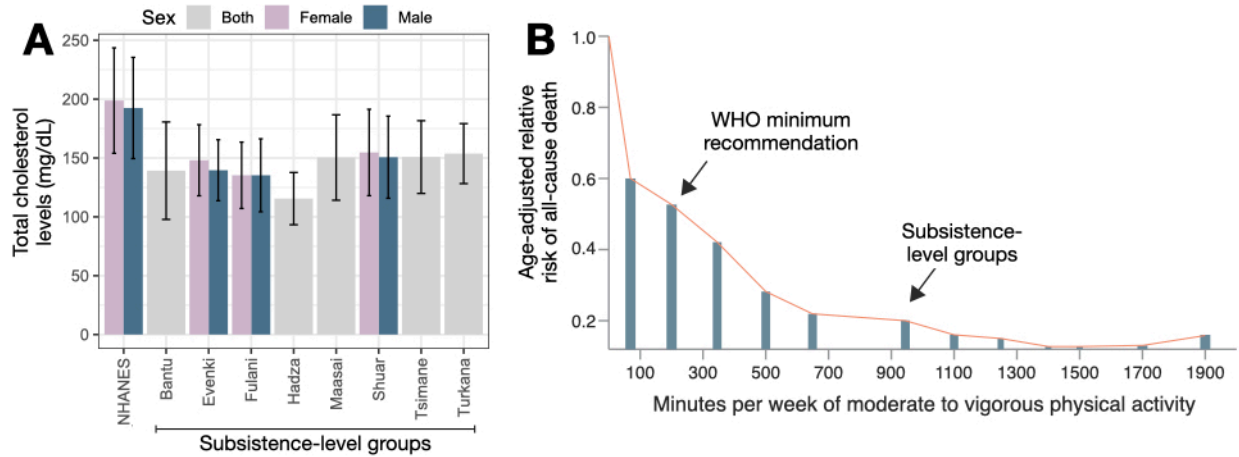

**Figure 4. Evidence for evolutionary mismatch at the phenotypic level.** A) Mean levels of total cholesterol are much lower in select subsistence-level populations relative to US adults (>18 years old) profiled as part of the National Health and Nutrition Examination Survey (NHANES) (56) (subsistence-level data from (17)). B) Evidence that, within industrialized populations, individuals accruing daily physical activity similar to those of men and women in subsistence-level societies experience similarly low rates of CVD as well as all-cause mortality from NCDs. Dose response relationship between minutes/week of moderate to vigorous leisure time physical activity and age-adjusted relative risk of death from a sample of 661,137 adult Americans and Europeans (57). The arrow for physical activity estimates in subsistence-level groups is based on studies of the Hadza (estimated at x=944 minutes (35)) and the Tsimane (x=924 minutes (58)).

**Current evidence for evolutionary mismatch at the genetic level**

As mentioned above, to fulfill the third criteria for mismatch, we would need to identify a locus for which 1) there is evidence of past selection and 2) performance of at least one allele varies across environments and confers inflated risk of an NCD in the novel environment (see also Figure 1B and Box 1). One would think this would be easy to find, but in fact there are only a handful of clear cases, despite good evidence for the existence of GxE interactions in general (59–62). One clear example of mismatch involves variants in the *APOL1* gene, which provides resistance to trypanosome infections. Given the prevalence of trypanosomes across Africa, beneficial alleles are found at high frequency in African populations as well as African Americans. However, these same variants confer elevated kidney disease risk in African Americans living in the US (63, 64).

Another example is related to the "thrifty genotype" hypothesis (14), which suggests that individuals living in environments where food is unpredictably and periodically scarce should experience selection to store body fat in times of plenty. Recently, an intriguing variant was found in Samoans, who are also susceptible to extreme obesity when eating a Western diet: a single amino acid variant (p.Arg475Gln) in the *CREBRF* gene exhibits signatures of past selection and is currently associated with a 1.3-fold increased risk of obesity (though puzzlingly, also a 1.6-fold decreased risk of type 2 diabetes). Subsequent functional work in cell culture models demonstrated that p.Arg475Gln has direct effects on metabolism, reducing energy use while increasing lipid storage (65).

In addition to these well-characterized examples (see also Figure 2 of (66)), recent genomic work has shown that, in aggregate, variants that serve as modern-day risk alleles for particular NCDs (namely CVD and autoimmune diseases) are more likely to show signatures of past selection relative to non-risk alleles (67–69). More broadly, there is now ample evidence that human populations can adapt to their unique ecologies quite quickly (70), setting the stage for mismatches when local conditions shift. For example, within the last 10,000 years, the high *P. vivax* malaria risk experienced by West Africans has selected for changes to a key chemokine receptor encoded by the *DARC* gene (71, 72), while the spread of dairying in Europe has selected for lactase persistence through changes in the regulation of the *LCT* gene (73, 74). As pathogen environments and diets inevitably change, local adaptation sets the stage for mismatches to occur.

**Consensus recommendations for a new path forward: integrating genomic tools and partnerships with transitioning populations**

In principle, GxE interactions are most simply identifiable using a mismatch framework by testing for environmentally-dependent genetic effects in transitioning populations. However, in practice, this would be difficult because most NCDs arise from many small genetic effects distributed across the genome. Further, the standard approach to resolve this needle-in-a-haystack problem—using a massive sample size—is difficult in small-scale groups who typically have modest population sizes. Instead, we discuss how advanced genomic methods can be combined with the mismatch framework in a principled way to quantify the role of GxE interactions in NCDs in subsistence-level settings.

First, we can improve GxE test power in transitioning populations by focusing on genetic loci with already demonstrated evidence for phenotypic relevance, for example, 1) those with evidence for recent selection in the study group or 2) those that have already been discovered in urban/industrialized environments. For example, recent work on the *APOE* locus found that the *E4* variant—a well-known risk factor for CVD and Alzheimer's disease in HICs—is associated with lower innate inflammation and may have beneficial effects on lipid moderation and cognition in a high pathogen/low obesity environment (75–77). We might expect similar successes in elucidating GxE mismatches at other well-known risk loci that replicate across HICs (e.g., *FTO, ADCY3, BRCA1/2*). A related approach is to test for GxE enrichment at the level of known genes or pathways, generalizing single SNP tests. These set-based approaches (i.e., that target predefined sets of loci) may also perform well in transitioning populations, even if the specific causal variants are not shared.

Second, polygenic approaches that integrate GxE signals across the genome can improve power when studying complex traits like NCDs. For example, recent methodological developments have extended the popular polygenic risk score (PRS) framework to allow for PRS-environment interaction tests, thus providing a polygenic GxE test (78–80). This approach has so far been used to show how diet and other lifestyle factors modulate the genetic risk of obesity (81–83). While polygenic approaches such as PRS sacrifice variant-level resolution, they yield much greater power to detect GxE interactions, an invaluable exchange for quantifying evolutionary mismatch in transitioning populations. Three downsides however are that: 1) compared to single, large-effect allele results, one can be left with no suggestion of underlying mechanism; 2) for PRS-environment interaction tests, power unavoidably depends on the predictive power of the PRS as well as its portability across contexts and ancestries, which is a clear problem given that most PRS work has focused on European ancestry individuals in HICs; and 3) again for PRS-environment interaction tests, an underlying assumption is that risk effects are systematically stronger in one environment than another (84).

Finally, we can add power and interpretability for GxE interactions using intermediate phenotypes like gene expression, DNA methylation, and chromatin accessibility. One approach is to impute these functional genomic features from genotype data and then test them for environmental interaction (e.g., akin to a GxE version of transcriptome-wide association studies (TWAS) (85, 86). The imputation step can use large, publicly available functional genomic datasets from HICs, but will improve when similar datasets are available for the study populations. A second approach is to test GxE in the map from genotype to functional genomic feature by identifying environmentally-sensitive variants that impact nearby gene expression, DNA methylation, chromatin accessibility, etc; this "molecular QTL" framework has so far proven very powerful and could be extended to transitioning populations (59, 87, 88). Moreover, GxE molecular QTLs can be validated experimentally by exposing cell lines or model organisms to stimuli that mimic aspects of the environmental gradients experienced by transitioning populations; indeed, this can pinpoint key components of the incredibly complex environmental shifts that drive GxE. Finally, a third option is to use functional genomic experiments to narrow the search space, by first identifying regulatory elements that respond to mismatch-relevant environments. For example, Garske and colleagues recently identified chromatin elements that respond to dietary fatty acids in adipocytes and then focused GxE follow up work on variants in these responsive elements. By doing so, they were able to gain power to search for interaction effects between genotype and dietary saturated fat intake on BMI (89). Similar *in vitro* functional genomic experiments (using field-collected samples) could be leveraged to target regions of the genome that may be most important for responding to key aspects of lifestyle transitions.

**Payoffs for NCD prevention and treatment**

Testing the degree to which GxE interactions arise from evolutionary mismatch would answer mechanistic questions about how GxE interactions manifest. For example, are loci that were involved in adaptation to a population's past environment more likely to exhibit GxE effects when the environment shifts? To what degree does the nature of GxE interactions vary across ancestries with distinct evolutionary histories? What is the envelope of "optimal" human environmental conditions that do not provoke mismatch? Molecular insights into evolutionary mismatch would allow us to prioritize the study of genetic variants that may adversely affect health outcomes in novel environments. It would also enable prediction of potential future adverse environments that could accelerate the onset of disease, and it could help us refine explanations for already observed ancestry-related differences in disease susceptibility.

The studies we recommend would also advance our understanding of health issues in minority, indigenous, and other underrepresented groups. Most subsistence-level populations in low- and middle-income countries (LMICs) are facing rapid rises in NCD risk, and the limited reports from these counties suggest that population responses to urbanization and market-integration are highly variable. Studies of European ancestry individuals in HICs are not well-suited to explain why. Partnering with transitioning groups to conduct evolutionarily and culturally informed studies is needed to better serve their health concerns.

**Conclusions and future directions**

The basic argument of this review is that we can further our understanding of evolution as well as the genetic architecture of human disease by combining genomic tools with studies of transitioning populations (as has been discussed previously (16), though not in the context of genomics). This recommended path improves on current approaches, which typically rely on "brute forcing" GxE scans across many SNPs and many environments. Instead, we advocate for using evolutionary theory to parse *a priori* which G and E we expect to interact. Doing so would boost power, better position us to understand and predict GxE interactions in the etiology of NCDs, and provide much needed insight into urgent health issues affecting vulnerable populations around the world.

Because the interdisciplinary perspective we take here necessarily touches on several fields, we did not attempt an exhaustive review of research on either evolutionary mismatch or GxE interactions (instead, we refer readers to excellent existing work (6, 12, 13, 15, 93, 94)). However, there are several interesting new directions in these fields that are worth highlighting. For example, a growing body of work has begun to conceptualize the human microbiome as an evolved trait that is currently "mismatched" to its environment, often with serious health implications (95). Given that 1) the microbiome is under host genetic control and can therefore be a target of natural selection (96), and 2) industrialization can induce large scale changes in gut microbial communities (97–99), this is an exciting area in which to investigate GxE interactions that generate mismatch diseases. Another emerging research topic is sex differences in the response to lifestyle change: several recent studies have found that women experience greater NCD risk following economic and nutritional transitions than men (17, 24, 100, 101), yet how sex-specific genetic, physiological, or environmental variation interact to produce this phenomenon is still unknown. Finally, it is well-established that early life experiences are important for predicting NCD risk later in life (102–104), and the timing of lifestyle change, as well as the degree to which individuals experience environmental mismatches *within their lifetimes*, may therefore be important to consider and to intersect with GxE frameworks (Box 3). In many cases, long-term partnerships with focal communities have already led to the creation of longitudinal datasets well positioned to take a lifecourse approach. Moving forward, we expect that longitudinal perspectives on environmental change, NCD risk, and GxE interactions will be especially fruitful.

**Acknowledgments**

This work was supported by a postdoctoral research fellowship to AJL from the Helen Hay Whitney Foundation as well as grants from the Searle Scholars Program, Canadian Institute for Advanced Research, and the National Institutes of Health (R35-GM147267). This work was also supported by grants from the National Institutes of Health to JFA (R01-ES029929 and R35-GM124881). We thank all participants from the "Evolutionary Mismatch Hypothesis in the Genomics Era" symposium, which generated many of the ideas discussed here.

**Author contributions**

AJL and JFA conceived the idea for the review. All authors drafted and edited the review.

**Competing interests**

The authors declare no competing interests.

**Boxes**

**Box 1. GxE interactions in population genetics: definitions and related concepts**

In population genetics, the simplest conceptualization of a GxE interaction involves three genotypes for a single bi-allelic locus, with each of the three genotypes found in two different environments and with fitnesses varying across these six conditions (Figure 3C). At equilibrium, this population will harbor, among other types of genetic variation, 1) alleles that have been selected to high frequency as a consequence of positive selection (i.e., selection on a trait value in a particular direction) and 2) alleles that are at intermediate frequency as a consequence of stabilizing selection (i.e., selection to keep trait values near an optimum). Now let's suppose the environment changes quickly: previously selected alleles may now be associated with a trait that is no longer beneficial, and even disease-causing, but they will remain at high frequency for some time before selection is able to purge them. Note that loci with no genetic variation (e.g., fixed beneficial mutations) could still be involved in mismatches in the new environment, but in the absence of genetic variation we will be unable to identify them.

In addition to GxE interactions, another population genetic concept relating to evolutionary mismatch and the modern increase in NCDs is *decanalization (105)*. Canalization refers to the process of stabilizing selection that minimizes genetic variation associated with fitness-related traits in a given environment. Decanalization, then, is a perturbation from this state that reveals genetic variation for health- or disease-associated phenotypes (106). Though similar, evolutionary mismatch is more specific than decanalization. Evolutionary mismatch can occur without having a previously canalized trait, and is a more general term not necessarily linked to stabilizing selection. Decanalization is always a form of evolutionary mismatch, but not the other way around. A final term that is distinct from all of these is *robustness*. Robustness refers to a property of individual genotypes, wherein they are able to retain an advantageous phenotype despite genetic or environmental hazards. In contrast, evolutionary mismatch and decanalization are population-level phenomena.

**Box 2. Ethical considerations of conducting genomic work in diverse populations**

Community engagement and ethical research is fundamental to achieving the broader vision of this Consensus. There is widespread consensus that broader population representation in biomedical research is critical for reducing health disparities (107), but moving forward on this agenda requires that we simultaneously acknowledge and learn from past mistakes and abuses.

At the heart of ethical considerations in genetics research is a situation in which diverse populations are dually under-represented and under-consulted (108). Recent work has outlined best practices for overcoming these issues (108–115). For example, Claw et al. (109) suggest six principles of research ethics: 1) understand community sovereignty and research regulations; 2) engage and collaborate; 3) build cultural competencies; 4) improve transparency; 5) build local research capacity; and 6) disseminate research in accessible formats. The common thread behind these principles is the importance of building trustful and long-term relationships based on principles of dynamic consent, reciprocity, beneficence, and sovereignty. In our own experience, building these sorts of relationships takes time (typically years) but is essential to do before engaging in research.

Basic research with populations in LMICs can lead to important insights, yet the value-added benefits from basic research (e.g., shaping health policy based on epidemiological trends, and/or the development of novel treatment strategies) often can take decades to materialize. Mechanisms for participant community involvement in these longer-term benefits should be explicitly embedded in initial plans (107). It is also important to recognize that community benefits can extend beyond the research itself. The needs and desires of local communities will vary widely, but populations in LMICs may face problems that are deeply inter-connected and often stem from systemic discrimination: poor nutrition and sanitation (often due to environmental degradation), minimal access to education, few economic opportunities, and loss of land rights. The priorities of communities will seldom match perfectly with the aims of scientists, especially when participant communities lack basic infrastructure and face discrimination. Prioritizing solutions to these problems is an opportunity to have great impact that will require cooperation between researchers, study participants, universities, NGOs, governments, and funding bodies.

**Box 3. Life course perspectives on NCD risk**

Development is a period of heightened environmental sensitivity, and challenging experiences early in life increase lifelong risk of most NCDs (102, 104, 116). Subsistence-level societies are an under-utilized yet potentially powerful model for studying early life influences on NCD risk. Many of these groups are currently experiencing rapid lifestyle changes leading to 1) extreme variation in early life conditions within a single population and 2) frequent mismatch between early life and adult environments—a situation that is thought to put individuals at risk for later life health issues (117-119). Point #1 provides a clear opportunity to leverage the distributional extremes to study early life effects on health (25, 122). Further, point #2 affords us the opportunity to compare outcomes when individuals experience within-lifetime environmental "matches" versus "mismatches". To date, studies of industrial transitions have come to mixed conclusions about the importance of within-lifetime mismatches (17, 39, 123, 124). More work in this area is urgently needed to understand when, why, and how early life experiences shape adult health in these groups.

Genomic tools applied to populations undergoing lifestyle change could also provide valuable insight into *how* early life experiences become "embedded" into lifelong physiology. At the molecular level, this process is thought to be mediated by stable changes in gene regulation (e.g., DNA methylation, chromatin accessibility, and gene expression). However, many gene regulatory elements are also dynamic and responsive to environmental perturbations throughout life. This fact leads to challenges in disentangling the effects of early versus later life environments, especially when the two are highly correlated (as is common in HICs). In contrast, subsistence-level groups in transition often experience decoupled early life and adult experiences, which could be leveraged to disentangle early versus later life influences. Genotype data collected for the same individuals could also be used to identify rarely studied GxE interactions where the "E" encompasses early life experiences (125–127). Overall, integrative studies of transitioning populations are primed to reveal which individuals will be most susceptible to NCDs during lifestyle transitions as well as when in the life course these exposures matter most.